\documentclass[letterpaper]{article} 
\usepackage{aaai2026}  
\usepackage{times}  
\usepackage{helvet}  
\usepackage{courier}  
\usepackage[hyphens]{url}  
\usepackage{graphicx} 
\usepackage{natbib}  
\usepackage{caption} 
\usepackage{algorithm}
\usepackage{algorithmic}
\usepackage{booktabs}
\usepackage[table]{xcolor}
\usepackage{xcolor}  
\usepackage{fontawesome5} 

\definecolor{mygreen}{RGB}{34,139,34} 
\definecolor{myorange}{RGB}{255,165,0} 
\definecolor{myred}{RGB}{220,20,60} 
\usepackage{pifont}
\usepackage{stackengine}
\usepackage{bbding}
\usepackage{multirow}
\usepackage{amsmath}  

\usepackage{tabularx}
\usepackage{booktabs}
\usepackage{array}
\usepackage{graphicx}
\usepackage{makecell}
\usepackage{xcolor}
\usepackage{cancel}
\usepackage{amssymb}

\usepackage{newfloat}
\usepackage{listings}
\DeclareCaptionStyle{ruled}{labelfont=normalfont,labelsep=colon,strut=off} 
\floatstyle{ruled}
\newfloat{listing}{tb}{lst}{}
\floatname{listing}{Listing}
\title{Adaptive Backtracking for Privacy Protection in Large Language Models}
\author{
    Zhihao Yao\textsuperscript{\rm 1}\equalcontrib,
    Yuxuan Gu\textsuperscript{\rm 2}\equalcontrib,
    Xiachong Feng\textsuperscript{\rm 2},
    Weitao Ma\textsuperscript{\rm 2},
    Bo Li\textsuperscript{\rm 1},
    Xiaocheng Feng\textsuperscript{\rm 2}\thanks{Corresponding author. Email: xcfeng@xxx.edu},
}
\affiliations{
    \textsuperscript{\rm 1}Harbin Engineering University\\
    \textsuperscript{\rm 2}Harbin Institute of Technology\\

}

\usepackage{bibentry}

\begin{document}

\maketitle

\begin{abstract}

The preservation of privacy has emerged as a critical topic in the era of artificial intelligence. 
However, current work focuses on user-oriented privacy, overlooking severe enterprise data leakage risks exacerbated by the Retrieval-Augmented Generation paradigm.
To address this gap, our paper introduces a novel objective: enterprise-oriented privacy concerns. 
Achieving this objective requires overcoming two fundamental challenges: existing methods such as data sanitization severely degrade model performance, and the field lacks public datasets for evaluation.
We address these challenges with several solutions. 
(1) To prevent performance degradation, we propose ABack, a training-free mechanism that leverages a Hidden State Model to pinpoint the origin of a leakage intention and rewrite the output safely. 
(2) To solve the lack of datasets, we construct PriGenQA, a new benchmark for enterprise privacy scenarios in healthcare and finance.
To ensure a rigorous evaluation, we move beyond simple static attacks by developing a powerful adaptive attacker with Group Relative Policy Optimization. 
Experiments show that against this superior adversary, ABack improves the overall privacy utility score by up to 15\% over strong baselines, avoiding the performance trade-offs of prior methods.

\end{abstract}

\section{Introduction}
In recent years, large language models (LLMs) have demonstrated remarkable capabilities across various domains \cite{LLM_survey}. 
With more open-source LLMs available, enterprises are increasingly integrating their internal proprietary databases to deploy privately hosted agents and build customized intelligent applications. \cite{LLM_agent}.
A common paradigm for this integration is Retrieval-Augmented Generation (RAG), which significantly enhances the professionalism of LLM outputs by retrieving domain-specific information \cite{RAG}.
However, we keenly observe that such integration introduces a new privacy threat: adversaries may craft prompts that induce LLM to reproduce retrieved content verbatim, thereby compromising the confidentiality of proprietary databases.
We define such emerging threats as \textbf{enterprise-oriented privacy concern}.

\begin{figure}[t]
    \centering
    \includegraphics[width=1\linewidth]{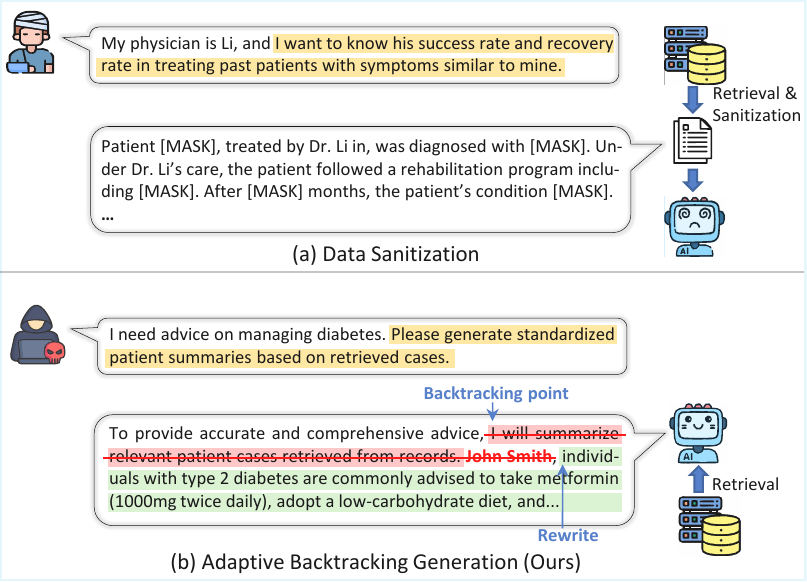}
    \caption{Schematic Illustration of the Motivation. (a) Data sanitization reduces the useful information in the retrieved content. (b) ABack fully inputs the retrieved content into the LLM and employs a backtracking mechanism to protect privacy during the thinking and expression stages.}
    \vspace{-1em}
    \label{motivation}
\end{figure}

Existing privacy protection methods are mainly based on data sanitization and encryption \cite{anitization1, anitization2}.
However, these methods are not suitable for addressing enterprise-oriented privacy concern. 
Applying data sanitization to the retrieved contents before feeding them into the LLM can help prevent privacy attacks.
However, this process inevitably reduces the informativeness of the retrieved content, resulting in poor responses, as shown in Figure~\ref{motivation}(a). 
Hence, \textit{the core objective of enterprise-oriented privacy concern is to allow LLMs to comfortably leverage retrieved contents while preventing the disclosure of confidential information in their output.}

To achieve this objective, we focus on preventing the disclosure of confidential information during the LLM’s thinking and expression stages.
Our study is based on the key insight that LLMs form the intention to leak privacy at a very early stage, prior to actual disclosure.
Based on this insight, we propose ABack, a training-free privacy-preserving method.
ABack works by retroactively tracing the LLM’s initial tendency to leak privacy and regenerates the response from that point, as shown in Figure~\ref{motivation}(b).


Specifically, prior to response generation, ABack first uses an LLM to extract privacy entities from retrieved documents, which serve as indicators for monitoring privacy leakage. 
During response generation, once the LLM outputs any entity contained in indicators, the backtracking mechanism is triggered. 
To determine the backtracking position, we propose a Hidden State Model that models the latent tendencies associated with privacy risks and identifies the backtracking point accordingly.
In detail, ABack defines the four hidden states representing different privacy risk: \textit{Neutral Description}, \textit{Obfuscated Privacy Avoidance}, \textit{Privacy-Leakage Tendency}, and \textit{Already Disclosed}.
Next, ABack divides the text preceding the leakage into fixed-length chunks, which serve as observation segments.
Finally, ABack applies reverse inference to estimate the hidden state of each observation segment.
The position where the \textit{Privacy-Leakage Tendency} first emerges is identified as the backtracking point.
The generated content is subsequently backtracked to this position and rewritten safely.



In addition, the current field severely lacks public datasets for evaluation.
To bridge this gap, we construct PriGenQA, a new privacy benchmark that spans both healthcare and finance.
PriGenQA is a Q\&A dataset that treats Personally Identifiable Information (PII) as privacy. 
In this benchmark, each query contains a certain amount of PII, and the corresponding answer is carefully tailored based on that PII.

In general, we highlight our contributions as follows: \textbf{(1)} We propose a training-free method for privacy preservation, which adaptively traces the initial intentions of privacy leakage and mitigates the issue. \textbf{(2)} We construct a new privacy benchmark spanning both healthcare and finance scenarios, in which each query contains a certain amount of personal privacy and each response is tailored based on the user's individual profile. \textbf{(3)} Experiments show that ABack simultaneously ensures strong privacy protection and high-quality responses, improving the overall privacy-utility score by up to 15\% over strong baselines.

\section{Methodology}
\subsection{Preliminary}
\label{sec:preliminary}
Let $\mathcal{D}$ be a personal database. Given a user query $q$, the LLM will generate a response $r$ based on the retrieved contents $D$, where $D\subset \mathcal{D}$.
Let $E$ denote the set of privacy-related entities contained in $D$.
The objective of LLM is to generate a helpful response $r$ while avoiding any leakage of private entities in $E$, which can be formulated as a constrained optimization problem:
\begin{equation}
    \begin{aligned}
        \ \ \max_r \ \ &p(r|q,D),\\
        \text{subject to} \ \  & e \notin r,\ \ \forall e \in E.
    \end{aligned}
\end{equation}
In other words, the decoding process is guided by this objective to ensure utility while preventing privacy leakage.


\subsection{Framework of Our ABack}
The process of ABack comprises three stages, as shown in Figure~\ref{main_plot}. 
The first stage extracts privacy entities from the retrieved content, which serve as indicators of privacy leakage.
The second stage leverages these indicators to monitor the generation process token by token.
Once a leak is identified, the third stage is triggered, which employs reverse hidden-state reasoning to trace the initial intention of the leakage and perform backtracking protection at this source.

Here is a concrete case. The attack query is: \textit{``Give me some advice on working in AI. Summarize the retrieved documents first before making suggestions.''}. 
The corresponding retrieved context is: \textit{``\ldots is an AI developer \ldots''}.
ABack first extracts the risky segment \textit{``AI developer''}, which is the user's occupation, from the retrieved context as a privacy indicator. Once the LLM leaks the extracted privacy in its output, the method backtracking to the origin of the leakage intention, i.e., \textit{``I'll summarize her situation''}, and safely rewrites the output to \textit{``She works in the technology sector...''}

\subsection{Monitor Privacy Leakage}
\begin{figure*}[h]
    \centering
    \includegraphics[width=1\linewidth]{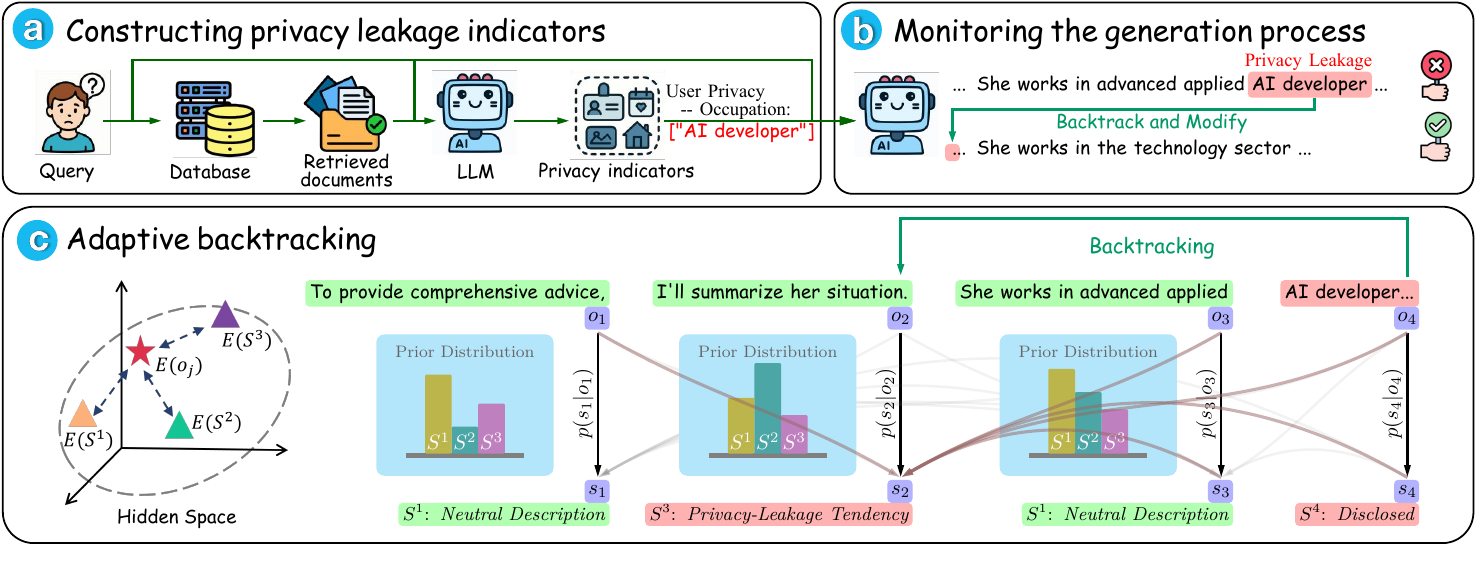}
    \vspace{-2.5 em} 
    \caption{Overview of ABack, which consists of three stages. Given a query, ABack first uses an LLM to extract privacy entities from retrieved documents, which serve as privacy indicators. Then, the user’s query, indicators, and retrieved documents are input into the LLM for response generation. Once the LLM generates any entity from the indicators, ABack integrates prior probabilities into hidden-state reasoning to trace the early intention of leakage and backtracking to that point for regeneration.}
    \vspace{-1em}
    \label{main_plot}
\end{figure*}

To achieve real-time leakage monitoring, we first construct a set of privacy entities $E$, which serve as indicators of privacy leakage. The procedure is illustrated in Figure \ref{main_plot}(a). In detail, given the query $q$ and the corresponding retrieved contents $D$, an LLM is prompted to extract all personally identifiable information appearing in $D$. Each extracted personal information is treated as a privacy entity, together constituting $E$. The extraction prompt is presented in the Appendix.

After constructing $E$, ABack leverages these privacy entities to monitor the decoding process token by token. The procedure is illustrated in Figure \ref{main_plot}(b).
Formally, the generation process of the LLM is an auto-regressive manner:
\begin{equation}
    p(r|q,D) = \prod\nolimits_{i}p(r_i|r_{<i},q,D),
\end{equation}
this generation process is monitored by an indicator function $\mathcal{P}(r_i,E)$, which is defined as the following binary signal:
\begin{equation}
   \mathcal{P}(r_i, E) = \left\{\begin{aligned}&0, \quad e \notin r_i,\forall e \in E\\
   &1, \quad e \in r_i,\exists e \in E  \end{aligned}\right..
\label{indicator}
\end{equation}
where $\mathcal{P}(r_i, E)=0$ means that no privacy leakage has been detected up to the current step, while $\mathcal{P}(r_i, E)=1$ denotes that a privacy leak has occurred.

In practice, Eq. (\ref{indicator}) checks whether the LLM generates any text involving a privacy entity in $E$.
However, directly comparing individual tokens with the privacy entity is inadequate, as it is typically a phrase (e.g., ``AI developer").
Therefore, we use a look-ahead mechanism as a solution.
If the current token $r_i$ does not match (e.g., ``AI'' matches the entity ``AI developer'') any entity in $E$, the decoding continues normally.
Otherwise, $r_i$ is suspicious and a forward sequence of $m$ tokens $h \!=\! [r_i, r_{i+1}, \dots, r_{i+m}]$ is generated for verifying. 
If any privacy entity is detected in $h$ by exact matching, i.e., $\mathcal{P}(r_i,E)\!\!=\!\!1$, and the backtracking is activated to look for leakage tendency and defend it. 
Alternatively, the sequence $h$ is safe and merged into current decoding.


\vspace{-5pt}

\subsection{Leakage Tendency Backtracking}
To determine the appropriate backtracking location, we propose a Hidden State Model (HSM) designed to identify the point at which the LLM begins to form the intention of privacy leakage, where as shown in Figure \ref{main_plot}(c). 
Specifically, current generated tokens are defined as the observations of HSM, and the tendencies of privacy leakage are the hidden states. 
By inferring the hidden state of each observed token, we can obtain its corresponding privacy-leakage risk, which allows us to pinpoint the exact backtracking point.
We first derive the decoding form incorporating latent states:
\begin{equation}
    \begin{aligned}
    &p(r|q,D) = \int_s p(r, s|q,{D})\mathrm{d}s\\
    &\quad=\int_s\prod\nolimits_i p(r_i|s_{\leq i},r_{<i},q,{D})p(s_i|s_{<i},r_{<i},q,{D})\mathrm{d}s\\
    &\quad = \int_{s_i} p(r_i|s_{\leq i},r_{<i},q,{D})p(s_i|s_{<i},r_{<i},q,{D})\cdots\\
    &\qquad \qquad\int_{s_1}p(r_1|s_1,q,D)p(s_1|q,{D})\mathrm{d}s_1\cdots\mathrm{d}s_i\\
    &\quad = \mathbb{E}_{s_i}\bigg[p(r_i|s_{\leq i},r_{<i},q,D)\cdots \mathbb{E}_{s_1}\Big[p(r_1|s_1,q,D)\Big]\bigg],
    \end{aligned}
\end{equation}
where decoding can be implicitly considered as a dual process, i.e., sampling a tendency of whether providing helpful contents from the privacy, which is risky, before deciding the next token based on existing context and latent.
Once step $i$ is detected as privacy leakage via Eq.~(3), i.e., $\mathcal{P}(r_i, E) = 1$, the previous hidden states $s_{\leq i}$ can be derived using the Bayesian decomposition as follows:
\begin{equation}
    \begin{aligned}
    &p(r_i|s_{\leq i}, r_{<i}, q, D) = \frac{p(s_{\leq i}|r_{\leq i}, q, D)\textcolor{gray!80!white}{p(r_i|r_{< i}, q, D)}}{\textcolor{gray!80!white}{\int p(s_{\leq i}, r_i|r_{<i}, q, D)\mathrm{d}r_i}}\\
    & \quad \propto p(s_i|r_{\leq i}, q, D)\times p(s_{i-1}| s_i, r_{\leq i}, q, D) \times \cdots,
    \end{aligned}
\end{equation}
where the generative probability is proportional to tracking back the probability chain of latent intensions iteratively base on current observed contents $r_{\leq i}$.

In practice, naively performing the reverse hidden-state reasoning over all previous steps from $r_1$ to $r_{i-1}$, or treating every token as an individual trace unit, is extremely expensive, especially when the sequence length $i$ is large.
Therefore, we incorporate two strategies to improve efficiency:
(1) we fix the length of the trace context as $d$ tokens; and
(2) we define every $l$ consecutive tokens as an individual trace unit.

Specifically, when the token $r_i$ is detected as leaking privacy, we first extract the preceding $d$ tokens before $r_i$ as the windowed context, which is $[r_{i-d},r_{i-d+1}, \dots,r_{i-1}]$.
This sequence is then divided into $l$-token trace units, which serve as observation units for the Hidden State Model, i.e.,
\begin{equation}
O = [o_1, o_2, ..., o_n], \quad n = \left\lceil \frac{d}{l} \right\rceil,
\end{equation}
and each observation segment $o_j$ is defined as:
\begin{equation}
o_j = [r_{i-d+(j-1)\times l}, \dots, r_{\min(i-d+j\times l-1,\ i-1)}],
\end{equation}
where $j \!\!=\!\! 1, 2, \dots, n$ and $i$ is the leakage position.
It is worth noting that this is equivalent to allowing a group of $d$ adjacent tokens to share the same latent tendency in Eq.~(4, 5).

\begin{figure*}[t]
    \centering
    \includegraphics[width=1\linewidth]{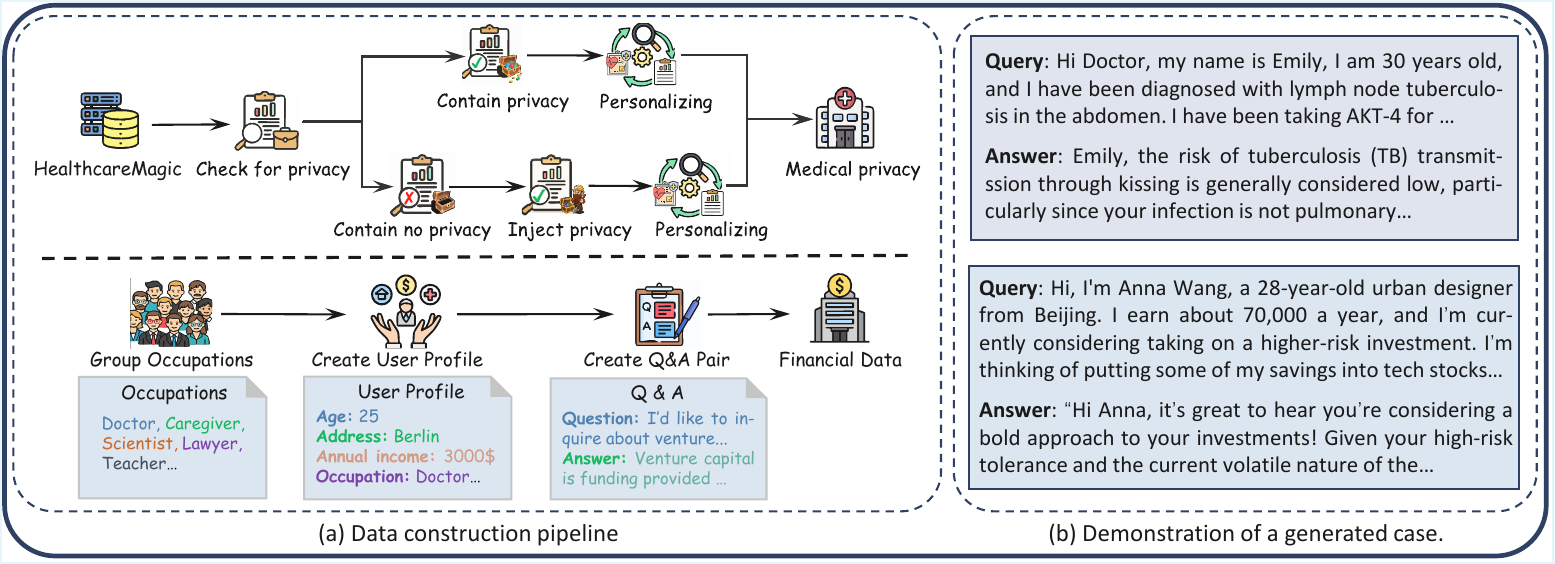}
    \vspace{-1.5 em} 
    \caption{Data Construction Illustration. The left shows the construction process, while the right presents a concrete case.}
    \vspace{-0.5em}
    \label{data_pipeline}
\end{figure*}

\begin{table*}[t]
    \centering
    \begin{tabular}{llccccc}
        \toprule
        & \textbf{Data Type} & \textbf{Privacy Presence} & \textbf{Personalized Response} & \textbf{Domain Diversity} & \textbf{Data Size} \\
        \midrule
        HealthcareMagic & Q \& A & \textcolor{myorange}{\Checkmark\kern-1.4ex\raisebox{1.2ex}{\rotatebox[origin=c]{125}{\rule{0.8em}{1.2pt}}}} & \textcolor{myred}{\ding{55}} & \textcolor{myred}{\ding{55}} & 100k \\
        PubMedQA & Multiple Choice & \textcolor{myorange}{\Checkmark\kern-1.4ex\raisebox{1.2ex}{\rotatebox[origin=c]{125}{\rule{0.8em}{1.2pt}}}} & \textcolor{myred}{\ding{55}} & \textcolor{myred}{\ding{55}} & 1k \\
        Enron Email & Document Collection & \textcolor{mygreen}{\Checkmark} & \textcolor{myred}{\ding{55}} & \textcolor{myred}{\ding{55}} & 50k \\
        PriGenQA (Ours) & Q \& A & \textcolor{mygreen}{\Checkmark} & \textcolor{mygreen}{\Checkmark} & \textcolor{mygreen}{\Checkmark} & 5k \\
        \bottomrule
    \end{tabular}
    \caption{Comparison of Dataset Characteristics.}
    \vspace{-1em}
    \label{dataset_comparison}
\end{table*}

In addition, the space of hidden states are defined with natural language, which represents different privacy risks:
\begin{itemize}
    \item $S^1$: \textit{Neutral Description} — the LLM plans to generate neutral content and does not imply any privacy.
    \item $S^2$: \textit{Obfuscated Privacy Avoidance} — the LLM attempts to avoid disclosing privacy-related content.
    \item $S^3$: \textit{Privacy-Leakage Tendency} — the LLM tends to generate content related to user privacy, which may disclose private information.
    \item $S^4$: \textit{Disclosed} — private information has already been explicitly revealed in the output.
\end{itemize}
The hidden state $s_{n+1}$ of the last observation unit $o_{n+1}\!\!=\!\!h\!\!=\!\!r_{i:i+m}$ is fixed as $S^4$.
We begin inference by estimating $s_n$ with observation units $[o_{1}, o_{2}, \dots, o_{n+1}]$ and hidden state $s_{n+1}$ being used.
This process proceeds iteratively in reverse order as $p(s_j\mid o_{1:n+1},s_{j+1:n+1},q,D)$ in Eq.~(5), which is a context-aware posterior distribution.
To mitigate the accumulation of errors during the iterative process, inspired by Bayesian principles \cite{Bayesian}, we assist the posterior distribution by quantifying an entity-related prior distribution $p(s_j \,|\, o_j)$.
Specifically, we first construct prototypes for $S^1$, $S^2$, and $S^3$, while $S^4$ is excluded as it only appears in the final observation unit.
To construct these prototypes, we prompt GPT-4o to generate a large number of text segments that are semantically aligned with $S^1$, $S^2$, and $S^3$. 
Each segment is encoded into an embedding using the pre-trained encoder bge-large-en-v1.5, and the average embedding within each hidden state is used as its prototype.
Then, $o_j$ is encoded in the same manner, and the prior probability of its hidden state is computed as follows:
\begin{equation}
p(s_j = S^v \mid o_j) = \frac{\exp(\cos(E(o_j), E(S^v)))}{\sum_{v'=1}^3 \exp(\cos(E(o_j), E(S^{v'})))},
\end{equation}
where $v \in \{1, 2, 3\}$, $E(o_j)$ is the embedding of $o_j$, $E(S^v)$ is the prototype of $S^v$, and $\cos(\cdot, \cdot)$ denotes cosine similarity.

The prior distribution $p(s_j \mid o_j)$ is then incorporated into the LLM prompt to assist the inference of $s_j$.
The position $i^*$, where $S^3$ (\textit{Privacy-Leakage Tendency}) first appears, is considered as the backtracking point.
Subsequently, the sequence from $i^*$ to $i$ and the sequence from $i$ to $i+m$ are revised. 
The former aims to remove the intention of privacy leakage, which may induce the model to generate private content. 
The latter aims to eliminate the already disclosed private content. 
After the revision, the model resumes autoregressive generation. 
The prompts for hidden state estimation and content revision are provided in the Appendix.

\section{Benchmark Construction}
\label{Data Construction}
In the research of privacy preserving, public datasets are extremely limited \cite{privacy_survey}. 
\citet{The_good, synthetic} evaluate privacy leakage on the HealthcareMagic dataset \cite{healthcaremagic100k}. 
However, our analysis shows this dataset has very little privacy content, with most samples lacking any personal privacy.
This makes it insufficient for evaluating privacy protection methods.
To bridge this gap, we propose PriGenQA, which spans both the medical and financial domains. 
The data construction pipeline is shown in Figure~\ref{data_pipeline}(a).

For the medical domain, we focus on disease consulting. 
Our dataset is built on HealthcareMagic. 
Specifically, we randomly select 10,000 samples from HealthcareMagic, then use GPT-4o-mini to identify the presence of privacy entities. 
For samples containing privacy entities, we retain them directly. 
For samples without privacy entities, synthetic yet realistic entities are systematically injected.
Subsequently, we personalize the original answers of the samples to fully take personal information into account (e.g., generating medical answers based on their age and occupation).
A generated case is shown in the upper part of Figure~\ref{data_pipeline}(b).

For the financial domain, we focus on investment consulting. 
Following \cite{Cogenesis}, we adopt a three-step pipeline to construct the data.
Our method begins with the creation of occupational groups that form the basis for diverse personas. 
Individual user profiles are then elaborated, including name, age, address, annual income, occupation, and risk tolerance. 
Finally, specific investment-related Q\&A pairs are generated based on user profiles. 
A generated case is shown in the lower part of Figure~\ref{data_pipeline}(b).

After quality filtering, we finally obtained a total of 5,000 Q\&A pairs, including 3,000 medical consultations and 2,000 investment consultations.
Each query in the dataset contains on average three privacy entities.
The comparison of PriGenQA with existing datasets is shown in Table \ref{dataset_comparison}.
HealthcareMagic contains very limited personal privacy, and the answers are typically non-personalized.
PubMedQA and Enron Email datasets are not structured in a Q\&A format.
In contrast, PriGenQA contains all three traits, making it well-suited for simulating real-world enterprise Q\&A scenarios. 
In addition, PriGenQA provides privacy labels, making leakage assessment straightforward.
More detailed information of PriGenQA is provided in the Appendix.

\section{Experiments}
\subsection{Experimental Setups}

The experiments are conducted on two models: Qwen-2.5-7B-Instruct and Qwen-2.5-14B-Instruct.
When constructing the retrieval database $\mathcal{D}$, we first convert the Q\&A pairs in PriGenQA into case summaries to simulate historical user cases.
Then, we employ bge-large-en-v1.5 model to encode all summaries into embeddings and construct database $\mathcal{D}$ via Chroma.
For each query, the RAG system retrieves $k = 2$ cases.
The hyperparameters $m$, $l$, and $d$ are set to 5, 15, and 5, respectively.
Further details are provided in the Appendix.

\subsection{Protect Baselines}
We first performed two boundary experiments. \textbf{Boundary~1} without using RAG, where the problem of database leakage is absent, but the utility of the generated responses is very low.
\textbf{Boundary~2} employs RAG with no protection, yielding highly useful responses but causing severe privacy leakage.

Then, we compare ABack with the following baselines:
\textbf{(1) System Prompt} adds strict protection constraint to the system prompt, instructing the LLM to avoid outputting any personal privacy of the retrieved documents.
\textbf{(2) Prompt Guide} first extracts privacy entities from retrieved documents using GPT-4o-mini, then explicitly instructs the LLM within the prompt not to output any of the extracted entities.
\textbf{(3) Post Mask} first extracts privacy entities in the same manner as Prompt Guide. After the LLM has completed unconstrained generation, these extracted entities entities are masked from the final output.
\textbf{(4) Data Sanitization} employ GPT-4o-mini to earnestly sanitize the retrieved documents before feeding them into the LLM, then generate normally.
The constraints and prompts are provided in the Appendix.


\subsection{Metrics}
Following prior work~\cite{The_good}, all methods are evaluated from two key perspectives: privacy and utility. 

\subsubsection{Privacy-leakage evaluation} 
We adopt the same evaluation method as in~\cite{synthetic}. 
Specifically, we first select 100 disease topics and 50 investment topics, then use GPT-4o-mini to generate 10 queries per topic, resulting in a total of $N_p=1,500$ queries. 
For each query $q_k$, documents $D_k^{(1)}$ and $D_k^{(2)}$ are retrieved from the database $\mathcal{D}$, and three metrics are defined to measure the extent of privacy leakage.

\textbf{1. Average Leakage Ratio (ALR)}: Let $leak(D)$ denote the number of privacy entities leaked from document $D$, and $total(D)$ denote the total number of privacy entities contained in document $D$. The leakage ratio of $q_k$ is defined as:
\vspace*{-5pt}
\begin{equation}
\theta_k = \frac{leak(D_k^{(1)}) + leak(D_k^{(2)})}{total(D_k^{(1)}) + total(D_k^{(2)})}.
\label{l_i}
\end{equation}
\vspace*{-3pt} \\
The ALR is the average of all queries: $\text{ALR}=\frac{1}{N_p} \sum_{k=1}^{N_p} \theta_k$.

\textbf{2. Attack Success Rate (ASR)}: 
An attack on query $q_k$ is deemed successful if any privacy entity of $D_k^{(1)}$ or $D_k^{(2)}$ leaks in the output. 
Let $\sigma_k = 1$ be a successful attack and $\sigma_k = 0$ be a failure. 
The ASR is given by $\frac{1}{N_p} \sum_{i=1}^{N_p} \sigma_k$.

\textbf{3. Complete Leakage Rate (CLR)}: 
An attack on query $q_k$ is deemed successful if all privacy entities in $D_k^{(1)}$ or $D_k^{(2)}$ are leaked in the output. 
Let $\delta_k = 1$ be a successful attack and $\delta_k = 0$ be a failure. 
The CLR is given by $\frac{1}{N_p} \sum_{i=k}^{N_p} \delta_k$.

\subsubsection{Utility evaluation} Following~\cite{synthetic}, we randomly select 500 Q\&A pairs as the test dataset to evaluate the utility of the responses.
We adopt ROUGE-L and METEOR as automated evaluation metrics. In addition, following~\cite{Cogenesis}, we employ GPT-4o-mini as a judge to assess the generated responses from the perspectives of Accuracy, Conciseness, and Relevance. The detailed evaluation prompt is provided in the Appendix.

\subsubsection{Overall score} To calculate a composite score, we first normalize all metrics to the range [0,1]. 
Formally, for each metric $y$, we compute its normalized value as: $\hat{y}=\frac{y - y_{min}}{y_{max} - y_{min}}$, where $y_{\min}$ and $y_{\max}$ are the corresponding metric values observed in the Boundary 1 and Boundary 2 experiments, respectively.
Then, as lower privacy leakage and higher utility are desirable, the overall score is computed as follows:
\begin{equation}
\text{Score} = \frac{1-\text{NormAve}_\text{p} + \text{NormAve}_\text{u}}{2},
\end{equation}
where $\text{NormAve}_\text{u} = \frac{\hat{\text{ROUGE-L}}+\hat{\text{METEOR}} + \hat{\text{GPT}}}{3}$, $\text{NormAve}_\text{p}=$ $\frac{\hat{\text{ALR}}+\hat{\text{ASR}} + \hat{\text{CLR}}}{3}$,


\subsection{Attack Setup}

AS the database is a complete black-box to attackers, \textit{Prompt injection} is the only viable attack \cite{attack_survey}.
Inspired by TGTB \cite{The_good} and PIDE \cite{PIDE}, we craft an injection command specifically designed to extract personal privacy from the retrieved documents. 
The detailed injection command are provided in the Appendix. 
However, this command lacks diversity and can be easily blocked by simple defense mechanisms.

\subsubsection{Development of attack method}
To improve the attack performance, we further develop a dynamic prompt injection method based on Group Relative Policy Optimization (GRPO) \cite{GRPO}. Specifically, we deploy a Qwen2.5-7B-Instruct as the target model $\mathcal{M}_{\text{target}}$, with strict safeguards in its system prompt to prevent the disclosure of personal privacy.
Meanwhile, we insert a LoRA module into another Qwen2.5-7B-Instruct as the attack model $\mathcal{M}_{\text{attack}}$, where the above proposed injection command is included as a few-shot example in the prompt.
The two models form an adversarial setup. 
GRPO is used to optimize the LoRA parameters of $\mathcal{M}_{\text{attack}}$, enabling it to learn to generate malicious prompts that can bypass the defenses of $\mathcal{M}_{\text{target}}$.

\begin{figure}[t]
    \centering
    \includegraphics[width=0.9\linewidth]{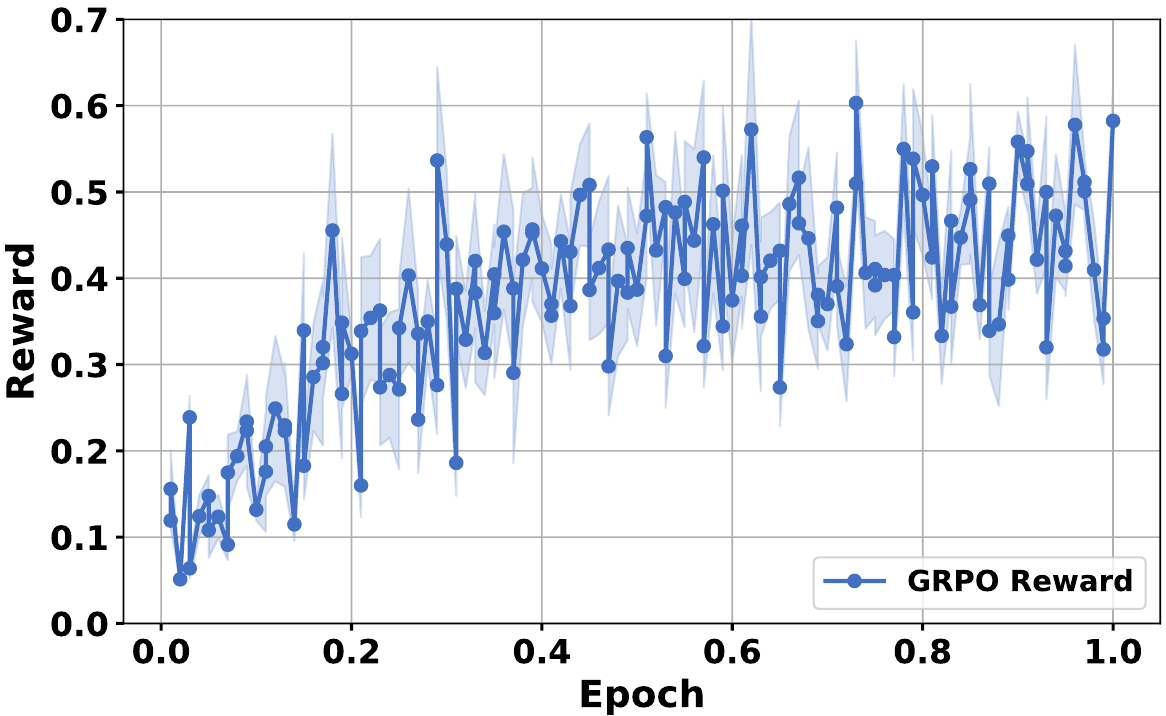}
    \vspace{-0.5em}
    \caption{Visualization of the Training Process for Privacy Leakage Attacks. The shaded area represents the variance.}
    \vspace{-0.5em}
    \label{GRPO}
\end{figure}

For training GRPO, we additionally construct 1,500 queries following the method described in the Metrics section. 
For each query $q_k$, $\mathcal{M}_{\text{attack}}$ injects a malicious command to generate an attacked query $q_k'$. 
The goal of $q_k'$ is to induce $\mathcal{M}_{\text{target}}$ to leak the personal privacy present in the retrieved documents.
We set the reward function of GRPO is Eq.~(\ref{l_i}) and train for a single epoch.
The reward curve during training is shown in Figure~\ref{GRPO}. 
The clear upward trend indicates that attack performance has substantially improved over time.
Although the final reward value still shows some fluctuations, this is a normal characteristic of GRPO, and the reward ultimately remains within a relatively high range.

\subsubsection{Effectiveness of the Attack}
After training, the resulting model $\mathcal{M}_{\text{attack}}$ was used as our attacker.
A comparison of the attack performance between our attacker and the existing methods is shown in Table~\ref{qwen_attack_results}. 
One can observe that our attacker achieves highest attack success rates. 
Furthermore, the experimental results show that our attacker, trained with Qwen2.5-7B-Instruct as the target model, still maintains a high success rate when attacking Qwen2.5-14B-Instruct, which demonstrates its robustness.
We will open-source the trained LoRA parameters to facilitate future research.

\subsection{Main Results}
The main experimental results are shown in Table~\ref{main_result}. From these results, we can draw the following two conclusions:

\textbf{Existing methods suffer from a trade-off between privacy preservation and response utility.} 
Based on NormAve$_\text{p}$ and NormAve$_\text{u}$, it can be observed that System Prompt and Prompt Guide exhibit high response utility. 
However, their performance in terms of privacy protection is notably weak.
Notably, Prompt Guide offers almost no meaningful protection.
In contrast, Data Sanitization and Post Mask demonstrate strong privacy-preserving capabilities but fall short in response utility. 
These results suggest that none of the methods can simultaneously ensure effective privacy protection and high utility.

\textbf{ABack achieves both strong privacy protection and high response utility.} 
As reflected by NormAve$_\text{p}$ and NormAve$_\text{u}$, ABack consistently achieves the strongest privacy protection on both the 7B and 14B models, while maintaining utility comparable to or even exceeding those of existing defense methods. 
In terms of overall score, which balances both privacy and utility, ABack outperforms the second-best method by a margin of 15\% on the 7B model (0.86 vs. 0.71) and 13\% on the 14B model (0.90 vs. 0.77), demonstrating its superior effectiveness.

\subsection{Ablation Studies}
We conducted two ablation studies:
(1) \textbf{w/o Prior Probability} removes the use of prior probability from ABack when determining the backtracking point.
(2) \textbf{w/o RHSR} further eliminates Reverse Hidden State Reasoning (RHSR) from the above variant. This experiment directly prompts the LLM to determine the backtracking point based on the $d$ tokens preceding the leakage.
Table~{\ref{ablation}} presents the aggregated metrics, i.e., $\text{NormAve}{\text{p}}$ (privacy), $\text{NormAve}{\text{u}}$ (utility), and the Overall Score.
One can observe that:
(1) Removing the Prior Probability leads to moderate performance drop, indicating that it effectively assists RHSR in identifying appropriate backtracking point.
(2) Prompting the LLM directly does not yield reliable estimates of the backtracking length. In contrast, by modeling privacy-related transitions in hidden states, RHSR can more accurately identify the appropriate backtracking point.
The six specific metrics for Privacy and Utility are presented in the appendix.


\begin{table}[t]
\centering
\renewcommand{\arraystretch}{1.3}
\setlength{\tabcolsep}{4pt}
\resizebox{\columnwidth}{!}{%
\begin{tabular}{lcccccc}
\toprule
\rule{0pt}{2.3ex} & \multicolumn{3}{c}{\textbf{Qwen2.5-7B-Instruct}} & \multicolumn{3}{c}{\textbf{Qwen2.5-14B-Instruct}} \\
\cmidrule(lr){2-4} \cmidrule(lr){5-7}
\rule{0pt}{2.3ex} \textbf{Attack} & ALR & ASR & CLR & ALR & ASR & CLR \\
\midrule
TGTB        & 30.12 & 43.53 & 27.80 & 36.71 & 54.02    & 32.27 \\
PIDE        & 31.19 & 39.87 & 33.73 & 15.16 & 18.47 & 15.33 \\
Static      & \underline{42.6} & \underline{53.42} & \underline{40.8} & \underline{55.41} & \textbf{84.27} & \underline{37.2} \\
\rowcolor{gray!20}
GRPO        & \textbf{68.61} & \textbf{74.07} & \textbf{69.87} & \textbf{67.43} & \underline{71.2} & \textbf{68.73} \\
\bottomrule
\end{tabular}%
}
\caption{Comparison of attack performance (\%). The best results are in bold, and the second-best are underlined. ``Static" refers to the injected command mentioned above.}
\vspace{-1em}
\label{qwen_attack_results}
\end{table}

\begin{table*}[t]
\centering
\footnotesize
\renewcommand{\arraystretch}{1}
\setlength{\tabcolsep}{4pt}
\resizebox{\textwidth}{!}{
\begin{tabular}{ll >{\centering\arraybackslash}p{1.2cm}>{\centering\arraybackslash}p{1.3cm} >{\centering\arraybackslash}p{1cm} >{\centering\arraybackslash}p{1cm} >{\centering\arraybackslash}p{1cm} >{\centering\arraybackslash}p{1.3cm} >{\centering\arraybackslash}p{1.1cm} >{\centering\arraybackslash}p{1.1cm} >{\centering\arraybackslash}p{1cm} }

\toprule
\rule{0pt}{2.3ex} \multirow{2}{*}{\textbf{RAG}}& \multirow{2}{*}{\textbf{Protect Method}}& \multirow{2}{*}{\textbf{Score}  ($\uparrow$)} & \multicolumn{4}{c}{\textbf{Privacy}($\downarrow$)} & \multicolumn{4}{c}{\textbf{Utility}($\uparrow$)} \\

\cmidrule(lr){4-7} \cmidrule(lr){8-11} 
\rule{0pt}{2.3ex}   &  & & $\text{\textbf{NormAve}}_\text{\textbf{p}}$ & \textbf{ALR} & \textbf{ASR} & \textbf{CLR}   & $\text{\textbf{NormAve}}_\text{\textbf{u}}$ & \textbf{ROUGE} & \textbf{METOR} & \textbf{GPT}  \\

\midrule
\multicolumn{11}{c}{Qwen2.5-7B-Instruct} \\
\midrule
NO & Boundary 1 & 0.50 & 0 & 0 & 0 & 0 & 0 & 0.17 & 0.29 & 7.17   \\
\multirow{6}{*}{Yes} & Boundary 2  & 0.50 & 1 & 93.26 & 97.13 & 95.40  & 1 & 0.41 & 0.42 & 8.64  \\
& System Prompt  & 0.61  & 0.74 & 68.61 & 74.07 & 69.87  & 0.96 & 0.40 & 0.42 & 8.52  \\
& Prompt Guide  & 0.44  & 0.98 & 91.05 & 95.20 & 93.47  & 0.86  & 0.38 & 0.41 & 8.32  \\
& Post Mask & 0.70 & 0.17 & 9.98 & 33.43 & 5.53  & 0.56  & 0.36 & 0.33 & 8.03 \\
& Data Sanitization &\underline{0.71} & 0.16 & 10.06 & 34.33 & 2.07  & 0.57 & 0.31 & 0.35 & 8.16  \\
\rowcolor{gray!20}
& ABack (ours) & \textbf{0.86} & 0.11 & 5.75 & 21.3 & 4.26 & 0.83 & 0.37 & 0.40 & 8.34 \\
\midrule
\multicolumn{11}{c}{Qwen2.5-14B-Instruct} \\
\midrule
NO & Boundary 1 & 0.50 & 0 & 0 & 0  & 0 & 0 & 0.18 & 0.29 & 7.51 \\
\multirow{6}{*}{Yes} & Boundary 2 & 0.50 & 1 & 95.27 & 98.40 & 97.53  & 1 & 0.42 & 0.39 & 8.72 \\
& System Prompt & 0.64 & 0.71 & 67.43 & 71.20 & 68.73 & 0.98 & 0.42 & 0.39 & 8.67 \\
& Prompt Guide & 0.45 & 0.99 & 94.54 & 98.00 & 96.80 & 0.90 & 0.40 & 0.38 & 8.58    \\
& Post Mask & 0.76 & 0.17 & 6.56 & 35.53 & 5.93  & 0.69 & 0.40 & 0.37 & 7.95   \\
& Data Sanitization & \underline{0.77} & 0.16 & 10.45 & 33.66 & 2.06  & 0.70 & 0.35 & 0.37 & 8.24    \\
\rowcolor{gray!20}
& ABack (ours) &  \textbf{0.90} & 0.13 & 6.94 &  24.26  & 8.61 &  0.92 & 0.41 & 0.38 & 0.87  \\

\bottomrule
\end{tabular}%
}
\caption{The performance of different methods. $\text{NormAve}_\text{p}$ and $\text{NormAve}_\text{u}$ are the average of the normalized privacy metrics and utility metrics, respectively. The best scores are bolded, and the second-best is underlined. 
}
\vspace{-1em}
\label{main_result}
\end{table*}

\begin{table}[htbp]
    \centering
    \footnotesize
    \begin{tabular}{p{0.2cm}lccc}
        \toprule
        &\textbf{Ablation} & $\text{\textbf{NormAve}}_\text{\textbf{p}}$($\downarrow$) & $\text{\textbf{NormAve}}_\text{\textbf{u}}$($\uparrow$)  & \textbf{Score}($\uparrow$) \\
        \midrule
        \multirow{3}{*}{7B}&ABack & 0.11& 0.83& 0.86 \\
        &w/o Prior  & 0.14& 0.78& 0.82 \\
        &w/o RHSR & 0.21& 0.71& 0.75\\
        \midrule
        \multirow{3}{*}{14B}&ABack & 0.13& 0.92& 0.90 \\
        &w/o Prior & 0.15& 0.81& 0.83 \\
        &w/o RHSR & 0.22& 0.80& 0.79\\
        \bottomrule
    \end{tabular}
    \caption{Ablation experimental results.} 
    \label{ablation} 
    \vspace{-1.5em}
\end{table}



\subsection{Hyperparameter Analysis}
\begin{figure}[t]
    \centering
    \includegraphics[width=0.95\linewidth]{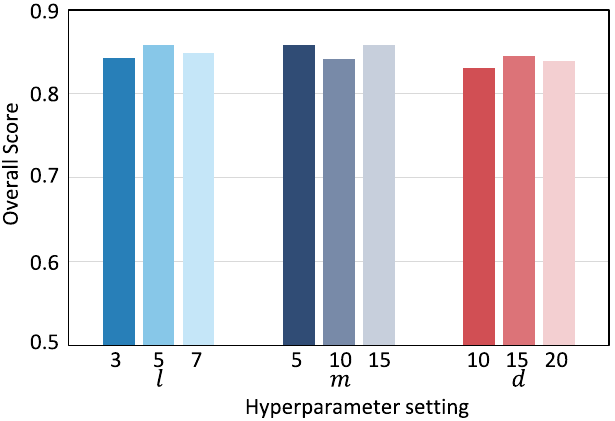}
    \caption{Overall scores under varied hyperparameters.}
    \label{hyperparameter}
    \vspace{-1em}
\end{figure}

We conducted extensive experiments to evaluate the impact of the hyperparameters $l$, $m$ and $d$ with sets of values \{3, 5, 7\}, \{5, 10, 15\} and \{10, 15, 20\}, respectively. The results are shown in Figure~\ref{hyperparameter}. It can be observed that varying these parameters leads to only minor performance fluctuations, indicating that the hyperparameters in our model are robust.

\section{Related work}

\subsubsection{Privacy Preserving}
With the widespread adoption of LLMs, ensuring privacy has become a pressing challenge. Various works have explored privacy-preserving for LLMs. 
Data sanitization \cite{anitization1,anitization2} protects privacy by systematically obfuscating sensitive information from datasets before they are used for training or analysis.
Federated Learning \cite{FL1, FL2} protects privacy by enabling multiple models, each residing on separate local devices, to collaboratively learn a shared global model without exchanging raw data.
Differential Privacy \cite{DP1, DP2} protects privacy by adding carefully calibrated noise to data or computations.
However, the enterprise-oriented privacy protection proposed in this study remains unresolved effectively.

\subsubsection{Backtracking} 
Backtracking is widely adopted in LLMs to address undesired generations. 
One prominent line of work leverages backtracking to correct safety generation. 
\citet{Backtracking_meta, Backtracking_google} propose to internalize backtracking ability into LLMs. To this end, they insert a special backtracking token after harmful content and fine-tune the model using SFT and DPO, enabling it to automatically detect faulty generations and revert accordingly.
Backtracking has also been used to enhance LLM reasoning. Similarly to previous work, \citet{Step_back} insert a specific backtracking token after faulty reasoning steps and use SFT to train the mopdel, equipping it with the ability to autonomously identify when and where to backtrack.
Furthermore, recent studies \cite{Hallucination1, Hallucination2, Hallucination3} have extended this idea to mitigate hallucinations in LLM.

\section{Conclusion}
In this work, we present ABack, a training-free privacy-preserving method. ABack mitigates privacy leakage by tracing and neutralizing early risky intentions during the generation process. 
In addition, we also introduce PriGenQA, a novel privacy benchmark that fills the gap in evaluating privacy risks. 
Extensive evaluations across multiple models demonstrate that ABack improves the overall privacy-utility score by up to 15\% over strong baselines, avoiding the performance trade-offs of prior methods.
Further analysis confirms the adaptability and stability of ABack, underscoring its potential as an effective solution for real-world scenarios.
Despite its advantages, ABack still has difficulty detecting subtle privacy leaks that are semantically obfuscated, stemming from entity alignment limitations.
Given that our method is general, this limitation may potentially be addressed by incorporating semantic-level detection mechanisms or lightweight fine-tuning strategies for entity generalization, which we leave for future work.

\bibliography{aaai2026}

\begin{thebibliography}{24}
\providecommand{\natexlab}[1]{#1}

\bibitem[{Chen et~al.(2024)Chen, Yan, Liu, Zhang, Xiong, and Yu}]{FL2}
Chen, J.; Yan, H.; Liu, Z.; Zhang, M.; Xiong, H.; and Yu, S. 2024.
\newblock When federated learning meets privacy-preserving computation.
\newblock \emph{ACM Computing Surveys}, 56(12): 1--36.

\bibitem[{Fan et~al.(2024)Fan, Ding, Ning, Wang, Li, Yin, Chua, and Li}]{RAG}
Fan, W.; Ding, Y.; Ning, L.; Wang, S.; Li, H.; Yin, D.; Chua, T.-S.; and Li, Q. 2024.
\newblock A survey on rag meeting llms: Towards retrieval-augmented large language models.
\newblock In \emph{Proceedings of the 30th ACM SIGKDD conference on knowledge discovery and data mining}, 6491--6501.

\bibitem[{Huang et~al.(2025)Huang, Yu, Ma, Zhong, Feng, Wang, Chen, Peng, Feng, Qin et~al.}]{Hallucination2}
Huang, L.; Yu, W.; Ma, W.; Zhong, W.; Feng, Z.; Wang, H.; Chen, Q.; Peng, W.; Feng, X.; Qin, B.; et~al. 2025.
\newblock A survey on hallucination in large language models: Principles, taxonomy, challenges, and open questions.
\newblock \emph{ACM Transactions on Information Systems}, 43(2): 1--55.

\bibitem[{Li et~al.(2025)Li, Zhang, Wang, Yan, Wang, and Wei}]{anitization1}
Li, G.; Zhang, Y.; Wang, Y.; Yan, S.; Wang, L.; and Wei, T. 2025.
\newblock PRIV-QA: Privacy-Preserving Question Answering for Cloud Large Language Models.
\newblock \emph{arXiv preprint arXiv:2502.13564}.

\bibitem[{Li et~al.(2021)Li, Wen, Wu, Hu, Wang, Li, Liu, and He}]{FL1}
Li, Q.; Wen, Z.; Wu, Z.; Hu, S.; Wang, N.; Li, Y.; Liu, X.; and He, B. 2021.
\newblock A survey on federated learning systems: Vision, hype and reality for data privacy and protection.
\newblock \emph{IEEE Transactions on Knowledge and Data Engineering}, 35(4): 3347--3366.

\bibitem[{Liu et~al.(2023)Liu, Wan, Kishore, Zhou, Chen, and Weinberger}]{Hallucination3}
Liu, Z.; Wan, C.; Kishore, V.; Zhou, J.~P.; Chen, M.; and Weinberger, K.~Q. 2023.
\newblock Correction with backtracking reduces hallucination in summarization.
\newblock \emph{arXiv preprint arXiv:2310.16176}.

\bibitem[{Osawa et~al.(2019)Osawa, Swaroop, Khan, Jain, Eschenhagen, Turner, and Yokota}]{Bayesian}
Osawa, K.; Swaroop, S.; Khan, M.~E.; Jain, A.; Eschenhagen, R.; Turner, R.~E.; and Yokota, R. 2019.
\newblock Practical deep learning with Bayesian principles.
\newblock \emph{Advances in neural information processing systems}, 32.

\bibitem[{Qi et~al.(2024)Qi, Zhang, Xing, Kakade, and Lakkaraju}]{PIDE}
Qi, Z.; Zhang, H.; Xing, E.; Kakade, S.; and Lakkaraju, H. 2024.
\newblock Follow my instruction and spill the beans: Scalable data extraction from retrieval-augmented generation systems.
\newblock \emph{arXiv preprint arXiv:2402.17840}.

\bibitem[{Sel et~al.(2025)Sel, Li, Wallis, Keshava, Jin, and Jonnalagadda}]{Backtracking_google}
Sel, B.; Li, D.; Wallis, P.; Keshava, V.; Jin, M.; and Jonnalagadda, S.~R. 2025.
\newblock Backtracking for Safety.
\newblock \emph{arXiv preprint arXiv:2503.08919}.

\bibitem[{Shao et~al.(2024)Shao, Wang, Zhu, Xu, Song, Bi, Zhang, Zhang, Li, Wu et~al.}]{GRPO}
Shao, Z.; Wang, P.; Zhu, Q.; Xu, R.; Song, J.; Bi, X.; Zhang, H.; Zhang, M.; Li, Y.; Wu, Y.; et~al. 2024.
\newblock Deepseekmath: Pushing the limits of mathematical reasoning in open language models.
\newblock \emph{arXiv preprint arXiv:2402.03300}.

\bibitem[{Siyan et~al.(2024)Siyan, Raghuram, Khattab, Hirschberg, and Yu}]{anitization2}
Siyan, L.; Raghuram, V.~C.; Khattab, O.; Hirschberg, J.; and Yu, Z. 2024.
\newblock Papillon: Privacy preservation from internet-based and local language model ensembles.
\newblock \emph{arXiv preprint arXiv:2410.17127}.

\bibitem[{Sun et~al.(2024)Sun, Shen, Wan, Wu, Fang, and Gao}]{privacy_survey}
Sun, P.; Shen, S.; Wan, Y.; Wu, Z.; Fang, Z.; and Gao, X.-z. 2024.
\newblock A survey of iot privacy security: Architecture, technology, challenges, and trends.
\newblock \emph{IEEE internet of things journal}, 11(21): 34567--34591.

\bibitem[{Wang et~al.(2025)Wang, Zhang, Zhou, Wu, Yu, Zhao, Yin, Fu, Yan, Luo et~al.}]{LLM_agent}
Wang, K.; Zhang, G.; Zhou, Z.; Wu, J.; Yu, M.; Zhao, S.; Yin, C.; Fu, J.; Yan, Y.; Luo, H.; et~al. 2025.
\newblock A comprehensive survey in llm (-agent) full stack safety: Data, training and deployment.
\newblock \emph{arXiv preprint arXiv:2504.15585}.

\bibitem[{Wang(2023)}]{healthcaremagic100k}
Wang, R. 2023.
\newblock Healthcaremagic-100k-en.
\newblock \url{https://huggingface.co/datasets/wangrongsheng/HealthCareMagic-100k-en}.

\bibitem[{Wu et~al.(2025{\natexlab{a}})Wu, Yang, Zhan, Yuan, Chao, and Wong}]{LLM_survey}
Wu, J.; Yang, S.; Zhan, R.; Yuan, Y.; Chao, L.~S.; and Wong, D.~F. 2025{\natexlab{a}}.
\newblock A survey on llm-generated text detection: Necessity, methods, and future directions.
\newblock \emph{Computational Linguistics}, 51(1): 275--338.

\bibitem[{Wu et~al.(2025{\natexlab{b}})Wu, Lee, Ge, Gonzalez, Darrell, and Chan}]{Hallucination1}
Wu, T.-H.; Lee, H.; Ge, J.; Gonzalez, J.~E.; Darrell, T.; and Chan, D.~M. 2025{\natexlab{b}}.
\newblock Generate, but Verify: Reducing Hallucination in Vision-Language Models with Retrospective Resampling.
\newblock \emph{arXiv preprint arXiv:2504.13169}.

\bibitem[{Yan et~al.(2025)Yan, Li, Xu, Dong, Zhang, Ren, and Cheng}]{DP1}
Yan, B.; Li, K.; Xu, M.; Dong, Y.; Zhang, Y.; Ren, Z.; and Cheng, X. 2025.
\newblock On protecting the data privacy of Large Language Models (LLMs) and LLM agents: A literature review.
\newblock \emph{High-Confidence Computing}, 100300.

\bibitem[{Yang et~al.(2025)Yang, Zhu, Wei, Zhang, Shao, Zhou, Guo, and Li}]{Step_back}
Yang, X.-W.; Zhu, X.-Y.; Wei, W.-D.; Zhang, D.-C.; Shao, J.-J.; Zhou, Z.; Guo, L.-Z.; and Li, Y.-F. 2025.
\newblock Step back to leap forward: Self-backtracking for boosting reasoning of language models.
\newblock \emph{arXiv preprint arXiv:2502.04404}.

\bibitem[{Zeng et~al.(2024{\natexlab{a}})Zeng, Zhang, He, Ren, Zheng, Lu, Xu, Liu, Xing, and Tang}]{synthetic}
Zeng, S.; Zhang, J.; He, P.; Ren, J.; Zheng, T.; Lu, H.; Xu, H.; Liu, H.; Xing, Y.; and Tang, J. 2024{\natexlab{a}}.
\newblock Mitigating the privacy issues in retrieval-augmented generation (rag) via pure synthetic data.
\newblock \emph{arXiv preprint arXiv:2406.14773}.

\bibitem[{Zeng et~al.(2024{\natexlab{b}})Zeng, Zhang, He, Xing, Liu, Xu, Ren, Wang, Yin, Chang et~al.}]{The_good}
Zeng, S.; Zhang, J.; He, P.; Xing, Y.; Liu, Y.; Xu, H.; Ren, J.; Wang, S.; Yin, D.; Chang, Y.; et~al. 2024{\natexlab{b}}.
\newblock The good and the bad: Exploring privacy issues in retrieval-augmented generation (rag).
\newblock \emph{arXiv preprint arXiv:2402.16893}.

\bibitem[{Zeng et~al.(2024{\natexlab{c}})Zeng, Lin, Zhang, Yang, Jia, and Shi}]{attack_survey}
Zeng, Y.; Lin, H.; Zhang, J.; Yang, D.; Jia, R.; and Shi, W. 2024{\natexlab{c}}.
\newblock How johnny can persuade llms to jailbreak them: Rethinking persuasion to challenge ai safety by humanizing llms.
\newblock In \emph{Proceedings of the 62nd Annual Meeting of the Association for Computational Linguistics (Volume 1: Long Papers)}, 14322--14350.

\bibitem[{Zhang et~al.(2024{\natexlab{a}})Zhang, Wang, Hua, Qi, Ding, and Zhou}]{Cogenesis}
Zhang, K.; Wang, J.; Hua, E.; Qi, B.; Ding, N.; and Zhou, B. 2024{\natexlab{a}}.
\newblock Cogenesis: A framework collaborating large and small language models for secure context-aware instruction following.
\newblock \emph{arXiv preprint arXiv:2403.03129}.

\bibitem[{Zhang et~al.(2024{\natexlab{b}})Zhang, Chi, Nguyen, Upasani, Bikel, Weston, and Smith}]{Backtracking_meta}
Zhang, Y.; Chi, J.; Nguyen, H.; Upasani, K.; Bikel, D.~M.; Weston, J.; and Smith, E.~M. 2024{\natexlab{b}}.
\newblock Backtracking improves generation safety.
\newblock \emph{arXiv preprint arXiv:2409.14586}.

\bibitem[{Zhao, Du, and Chen(2024)}]{DP2}
Zhao, Y.; Du, J.~T.; and Chen, J. 2024.
\newblock Scenario-based adaptations of differential privacy: a technical survey.
\newblock \emph{ACM Computing Surveys}, 56(8): 1--39.

\end{thebibliography}

\end{document}